\documentclass[prx,twocolumn,english,superscriptaddress,floatfix,longbibliography]{revtex4-2}

\usepackage{ulem}
\usepackage[dvipsnames]{xcolor}
\usepackage{graphicx}
\usepackage{dcolumn}
\usepackage{bm}
\usepackage{amssymb}
\usepackage{amsmath}
\usepackage{physics}
\usepackage{float}
\usepackage{xcolor}

\begin{document}
\preprint{APS/123-QED}

\title{YSR Bond Qubit in a Double Quantum Dot with cQED Operation}

\affiliation{Departamento de F\'{i}sica Te\'{o}rica de la Materia Condensada, Universidad Aut\'{o}noma de Madrid, Madrid, Spain}
\affiliation{Condensed Matter Physics Center (IFIMAC), Universidad Aut\'{o}noma de Madrid, Madrid, Spain}

\author{G. O. Steffensen}
\affiliation{Departamento de F\'{i}sica Te\'{o}rica de la Materia Condensada, Universidad Aut\'{o}noma de Madrid, Madrid, Spain}
\affiliation{Condensed Matter Physics Center (IFIMAC), Universidad Aut\'{o}noma de Madrid, Madrid, Spain}
\affiliation{Instituto Nicol\'{a}s Cabrera, Universidad Aut\'{o}noma de Madrid, Madrid, Spain}
\affiliation{Instituto de Ciencia de Materiales de Madrid (ICMM), \\ Consejo Superior de Investigaciones Científicas (CSIC), \\ Sor Juana Inés de la Cruz 3, 28049 Madrid, Spain}
\author{A. Levy Yeyati}
\affiliation{Departamento de F\'{i}sica Te\'{o}rica de la Materia Condensada, Universidad Aut\'{o}noma de Madrid, Madrid, Spain}
\affiliation{Condensed Matter Physics Center (IFIMAC), Universidad Aut\'{o}noma de Madrid, Madrid, Spain}
\affiliation{Instituto Nicol\'{a}s Cabrera, Universidad Aut\'{o}noma de Madrid, Madrid, Spain}

\date{\today}

\begin{abstract}

Connecting two half-filled quantum dots to two superconducting leads induces a competition of bonds, with the dots forming either an interdot exchange bond or two individual Yu-Shiba-Rusinov (YSR) screening bonds with the leads. Defining a qubit using these singlet parity bonding states provides dot charge noise protection, attributed to the chargeless nature of the screening quasiparticles, and magnetic noise protection, as the bonds guard against magnetic polarization. In this paper, we propose embedding a Double Quantum Dot (DQD) Josephson junction in parallel with a transmon to enable circuit Quantum Electrodynamics (cQED) measurements and operation of a YSR bond qubit. We demonstrate that, under realistic parameters, two-tone spectroscopy of the DQD can be performed, revealing a significant parameter regime suitable for qubit operation. Additionally, coherent manipulations of the bond states can be achieved through dot gates, and single-shot readout is enabled by measurements of a capacitively coupled resonator. Finally, we analyze noise sources and estimate gate noise on couplings as the primary source of qubit decoherence. Since this qubit is protected against nuclear Overhauser fields and does not rely on spin-orbit interactions for operation, a broader range of material platforms becomes available compared to current Andreev spin qubits.

\end{abstract}

\maketitle

\section{Introduction}
Superconducting transmon qubits \cite{Koch2007Oct} and semiconducting spin qubits \cite{Loss1998Jan} constitute two of the most promising platforms for scaleable quantum computing. Spin qubits offer long coherences and relatively small on chip size, while transmons provide the ability to control, manipulate, and entangle qubits over large distances using circuit Quantum Electrodynamics (cQED) techniques. The Andreev spin qubit \cite{Padurariu2010Apr, Hays2021Jul} attempts to combine the best aspects of both approaches by utilizing the spin degree of freedom of an Andreev Bound State (ABS) localized in a Josephson junction as the computational subspace. In this setup, the state dependent current phase relation (CPR) allows for easy integration into existing superconducting circuits, and the size of the junction is comparable to spin-qubits. In a recent work \cite{Pita-Vidal2023May}, an electrostatically defined quantum dot on a hybrid semi-superconductor nanowire was employed to create an Andreev spin qubit. This design benefits from charge noise protection and strong coherent coupling due to the large charging energy and spin-orbit coupling respectively. Furthermore, the large degree of tunability allows for deterministic preparation of the computational basis, as a qubit state can be tuned to be the system groundstate. Despite these advantages the qubit, as of its spin nature, is sensitive to local magnetic fluctuations arising from the dynamics of spin-full nuclei in the substrate, which is conjectured to limit its coherence time.

\begin{figure}[t!]
\includegraphics[width=0.9\linewidth]{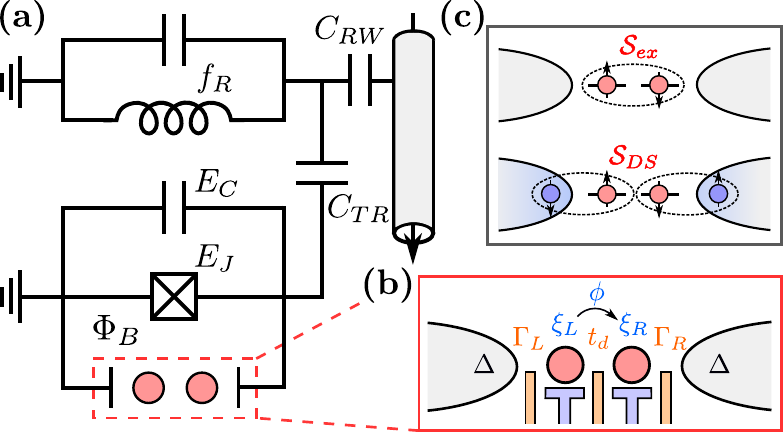}
\caption{\label{fig:principle}
(a) Schematic of a similar circuit as Ref.~\cite{Bargerbos2022Jul} with the DQD in parallel to a transmon ($E_J$ and $E_C$) and capacitively coupled to a resonator ($f_R$), which in turn is capacitively coupled to a waveguide. (b) Schematic of the DQD and gateable parameters. The qubit is protected from decoherence on parameters shown in blue, while susceptible to those in orange. (c) Illustration of the two states used as computational basis for the YSR bond qubit. Dashed circles indicate bonding interactions.
}
\end{figure}

In this paper we propose to embed a double quantum dot (DQD) in a similar circuit as in Refs.~\cite{Pita-Vidal2023May, Bargerbos2022Jul, Bargerbos2023Aug} composed of a parallel coupled transmon, a capacitively coupled resonator, and a waveguide as illustrated in Fig.~\ref{fig:principle} (a, b), to facilitate cQED manipulations and read-out. The coupling of a quantum dot to a superconductor generally leads to the formation of subgap states. Depending on the scale of dot charging energy, $U$, to superconducting gap, $\Delta$, these states will be more of an ABS ($\Delta > U$, \cite{Bauer2007Nov, MengFlorens2009}) or Yu-Shiba-Rusinov (YSR) screened ($\Delta \lesssim U$, \cite{Yu1965, Shiba1968, Rusinov1969}) character, with either screening quasiparticles residing in the lead (YSR), or an additional quasiparticle residing on the dot (ABS). We propose to instead of spin-states utilize a computational qubit basis composed of two singlet parity states; namely the exchange singlet, $\mathcal{S}_{ex}$, formed between the dots, and the doubly screened YSR singlet, $\mathcal{S}_{DS}$, where the spin on each dot is independently screened by the connecting lead \cite{EstradaSaldana2020Nov}. Both states are depicted in Fig.~\ref{fig:principle} (c) , with bonds indicating an exchange structure of the form,
\begin{align}
&\ket{\mathcal{S}_{ex}} = \frac{1}{\sqrt{2}}\left(\ket{\uparrow_L}\otimes\ket{\downarrow_{R}}-\ket{\downarrow_L}\otimes\ket{\uparrow_{R}}\right), \\
&\ket{\mathcal{S}_{DS}}= \ket{\mathcal{S}_L}\otimes\ket{\mathcal{S}_R}, \nonumber
\end{align}
with integer $L$ and $R$ denoting spin on left and right dot. The independent screened state on side $j$ is approximately given by,
\begin{equation}
\ket{\mathcal{S}_{j}} \approx \frac{1}{\sqrt{2}}\left(\ket{\uparrow_j}\otimes\ket{\downarrow_{QP,j}}-\ket{\downarrow_{j}}\otimes \ket{\uparrow_{QP,j}}\right), \label{eq:Sj}
\end{equation}
with $\ket{\uparrow_{QP,j}}$ indicating a state of screening quasiparticles in lead $j$, with total spin $1/2$ in the up direction~\cite{Bauer2007Nov, Kirsanskas2015, Baran2023Dec}.
Using this basis, quantum information is encoded in which bonds the dot electrons form, hence dubbed a \textit{YSR bond qubit}. In this setup we capture many of the same advantages as the quantum dot based Andreev spin qubit \cite{Pita-Vidal2023May}, such as dot charge noise protection, deterministic state preparation and cQED operation, but additionally we find quadratic protection from magnetic fluctuations, both local and global, and remove the requirement of spin-orbit coupling and magnetic Zeeman splitting. However, noise on tunnel couplings becomes the main source of qubit decoherence, as they set the energy of the bonding states, which we characterize and estimate. To illustrate experimental feasibility we choose similar circuit parameters as Ref.~\cite{Pita-Vidal2023May}, and utilize DQD parameters from a previously measured DQD superconducting junction \cite{EstradaSaldana2018Dec, EstradaSaldana2020Nov} which was also defined on a semi-super hybrid InAs/Al nanowire. Within these limitations we demonstrate that the qubit is both operational and measurable in a substantial range of parameters and with current material platforms.

In the following Section~\ref{sec:Setup} we introduce the DQD model and circuit used throughout the paper, then in Section~\ref{sec:Characterization} we show how the system can be characterized and an operational point for a qubit can be obtained. Section~\ref{sec:QubitOp} details operation and manipulation of the bond qubit, while noise and sweetspots are discussed in Section~\ref{sec:NoiseAndSweetspots}. Finally, in Section~\ref{sec:Discussion} we compare our proposal to other qubits before concluding. 

\section{Setup and Model}\label{sec:Setup}

\subsection{DQD Hamiltonian}
We model the serially coupled DQD as two Anderson impurities, each with a charging energy $U$ and coupled to a BCS lead with pairing $\Delta$ by a coupling $\Gamma_{L/R}$. The competition between the scales $U$, $\Delta$ and $\Gamma_{L/R}$ have theoretically been treated using various numerical methods and approximations \cite{Bergeret2006Oct, Lopez2007Jan, Karrasch2011Oct, Droste2012Sep, Rok2015, Pokorny2018May, Zonda2023Mar}. Choosing similar parameters as a previous DQD in a nanowire geometry \cite{EstradaSaldana2018Dec, EstradaSaldana2020Nov} we fix $\Delta=0.2$~meV, $U=5\Delta$, while $\Gamma_{L/R}$, dot energy, $\xi_{L/R}$, and interdot tunnel coupling, $t_d$, are considered tuneable. This places us in the regime of $U \gtrsim \Delta$ where YSR screening physics dominates, but induced superconductivity on the dots cannot be neglected \cite{Oguri2013}. As our goal is to obtain a tractable minimal qubit Hamiltonian we utilize the $\Delta\rightarrow \infty$ limit \cite{MengFlorens2009, Droste2012Sep}, for which the leads can be integrated out to yield an effective Hamiltonian,
\begin{align}
H_{DQD}(\phi) &= \sum_{j}\tilde{\xi}_j \left(n_j-1\right) + \sum_{j}\frac{\tilde{U}_j}{2}\left(n_j-1\right)^2 \nonumber \\
&+ \sum_j\left(\tilde{\Gamma}_j d^\dagger_{j\uparrow}d^\dagger_{j\downarrow}+\tilde{\Gamma}_j d_{j\downarrow}d_{j\uparrow}\right) \nonumber \\ 
&+ \sum_\sigma \left(\tilde{t}_d d^\dagger_{L\sigma}d_{R\sigma}+\tilde{t}_d^* d^\dagger_{R\sigma}d_{L\sigma}\right), \label{eq:FullHam}
\end{align}
with dot annihilation $d_{j\sigma}$ and number operators $n_j=\sum_\sigma d^\dagger_{j\sigma}d_{j\sigma}$, and the phase drop, $\phi$, occurring between the dots, $\tilde{t}_d=|\tilde{t}_d|e^{i\phi/2}$. Here $\phi$, which is the conjugate variable to the charge on $E_C$, is quantized and treated as an operator. In a recent paper~\cite{Zonda2023Mar}, it was demonstrated that $\Delta$ can be reintroduced to the $\Delta\rightarrow\infty$ limit by applying a parameter rescaling. The resulting model fits Numerical Renormalization Group results to a quantitative degree up to $U\sim 5\Delta$. This rescaling, dubbed the Modified Generalized Atomic Limit (MGAL), consists of,
\begin{align}
\tilde{\Gamma}_j &= \nu_j \Gamma_j, \hspace{0.2cm} \tilde{U}_j = \nu_j^2 U, \hspace{0.2cm} \tilde{\xi}_j = \nu_j^2 \xi_j\sqrt{1+\frac{2\Gamma_j}{\nu_j U}}, \nonumber \\
\tilde{t}_d &= \sqrt{\nu_L \nu_R}t_d, \hspace{0.3cm} \nu_j = \frac{1}{1+\Gamma_j/\Delta}. \label{eq:Rescallings}
\end{align}
While this approach accurately captures the subgap spectrum and CPR, it does not properly describe the wavefunction of the screened state. In this approach, as in the standard $\Delta\rightarrow \infty$ limit, the leads enter as effective pairings, $\Gamma_{L/R}$, on the dots, and for large pairing theindependent singlet groundstate on side $j$ becomes a superposition of even dot occupations, $\ket{\mathcal{S}_j} = u\ket{0} + v\ket{2}$. However, the actual wavefunction for our parameters would be more alike Eq.~(\ref{eq:Sj}), with a total spin-zero state on either side, corresponding to a superconducting equivalent of the Kondo groundstate~\cite{Bauer2007Nov}. With this in mind, we consider the Hamiltonian, and effective Hamiltonians derived from it, to be accurate in spectrum and CPR, but incomplete in calculating e.g. matrix elements by wavefunction overlap.

The Josephson energies of the DQD system is obtained from the eigenvalues, $E_n(\phi)$, of the $16$ dimensional Hamiltonian in Eq.~(\ref{eq:FullHam}), given by $V_n(\phi) = E_n(\phi) - E_n(0)$. The equilibrium Josephson energy is obtained from the free energy,
\begin{equation}
V_{eq}(\phi) = \sum_{n=1}^{16} \frac{V_n(\phi)\exp\left[-V_n(\phi)/k_BT\right]}{Z} \overset{T=0}{=} V_0(\phi),
\end{equation}
with $V_0(\phi)$ being the Josephson energy of the groundstate. In the following we set $T=0$, as temperatures small compared to the gap does not significantly modify the CPR.

\subsection{Microwave operation}
Following Refs.~\cite{Bargerbos2022Jul, Pita-Vidal2023May} we consider the DQD to be in parallel with a transmon characterized by a Josephson energy, $E_J/h = 13$~GHz and a capacitive energy, $E_C/h=0.2$~GHz, which yields a bare resonance frequency of $f_0 = 2\sqrt{2}\sqrt{E_C E_J}/h \approx 5$~GHz. Transfer of Cooper-pairs across the DQD is restricted by the large dot charging energies, and correspondingly for all DQD states $n$ the supercurrent $I_n(\phi) = \frac{2e}{\hbar}\partial_\phi V_n(\phi)  \ll I_c$ with $I_c = 2eE_J/\hbar$. Consequently, the application of a flux, $\Phi_B$, between the DQD and transmon Josephson junction exclusively tunes the static phase across the DQD, $\phi_B = 2\pi\Phi_B/\Phi_0$, with $\Phi_0 = h/2e$. In experiment $\Phi_B$ can originate either from a small magnetic field or from an inductively coupled wire. The joint circuit is thus characterized by the Hamiltonian,
\begin{equation}
H = 4 E_C \partial^2_\phi + E_J\left(1-\cos{\phi}\right) + H_{DQD}(\phi + \phi_B).
\end{equation}
This Hamiltonian couples DQD states to transmon states via the operator $\phi$, which in the transmon limit, $E_J \gg E_C$, can be expressed as $\phi \approx \phi_{0}(a+a^\dagger)$ with $\phi_0 = \left[2E_C/E_J\right]^{\frac{1}{4}}$ denoting the zero point phase fluctuation, and $a^\dagger$ the creation operator between $0$th and $1$st transmon level~\cite{Koch2007Oct}. Correspondingly, the Cooper-pair charge operator is given by $n_C = -i\partial_\phi \approx \frac{-i}{2\phi_0}(a-a^\dagger) $. Now, considering only terms from the two lowest transmon states we expand to lowest order in $|M_{nm}|\phi_0 \ll |E_n - E_m|$ and $|M_{nm}|\phi_0 \ll |E_n - E_m\pm hf_0|$ and project out excited states, yielding the transmon detuning stemming from $H_{DQD}$,
\begin{align}
h\delta f_{01,n} &= \phi_0^2 \frac{\partial^2 V_n}{\partial\phi^2} + 2\frac{|M_{nn}|^2}{hf_0} \label{eq:Detuning} \\ &+ 2\sum_{m\neq n} \left(\frac{|M_{nm}|^2}{E_m - E_n}-\frac{hf_0 |M_{nm}|^2}{(E_m-E_n)^2-(hf_0)^2}\right), \nonumber 
\end{align}
such that the total frequency is $f_{01,n} = f_0 + \delta f_{01,n}$ for DQD state $n$. The coupling element, 
\begin{equation}
M_{nm} = i|t_d|\phi_0\sum_\sigma\bra{n}d^{\dagger}_{L\sigma}d_{R\sigma}-d^\dagger_{R\sigma}d_{L\sigma}\ket{m}, \label{eq:MatrixElement}
\end{equation}
corresponds to an excitation or deexcitation of the transmon by the tunneling of an electron from dot-to-dot. We note that calculating $M_{nm}$ with state $n$ and $m$ obtained from MGAL, in general, leads to an overestimation of $M_{nm}$ for low energy states. This is compared to results of a zero-bandwidth (ZBW) approximation~\cite{Grove-Rasmussen2018Jun, Baran2023Dec} which more accurately captures the wavefunction. For details on this discrepancy see supplemental material~\cite{SM}.

The full transmon-DQD system is furthermore capacitively coupled to a superconducting resonator with a bare frequency $f_{R} = 6$~GHz. This results in a dispersive shift of the resonator frequency, $f_{R,n} = f_{R} + \delta f_{R,n}$, \cite{Blais2004Jun, Park2020Aug} with the primary contribution stemming from the lowest excitation of the transmon, 
\begin{equation}
\delta f_{R,n} \approx |g|^2\left(\frac{2}{f_{01,n}} - \frac{2 f_{01,n}}{f_{01,n}^2-f_R^2}\right).
\end{equation}
leading to the resonator frequency being dependent on the DQD state $n$. Here, the coupling $|g|$ is proportional to $C_{TR}$ as it is capacitively facilitated. Measurements can then be performed either via single-tone read-out of the resonator, which facilitates quadrature and single-shot measurements of the DQD, or two-tone spectroscopy. The latter is done by applying a frequency close to $f_R$ to the waveguide, and then via a second tone scan for $f_{01,n}$. On approaching $f_{01,n}$, this detunes the resonator out of resonance with $f_R$, yielding a clear signal. This measurement yields better resolution of the state $n$, as $\delta f_{01,n}$ is directly measured~\cite{Tosi2019Jan,Fatemi2022Nov,Matute-Canadas2022May, Bargerbos2022Jul, Pita-Vidal2023May}, but drives DQD transitions and therefore is not applicable for qubit read-out.

\begin{figure}[h]
\includegraphics[width=\linewidth]{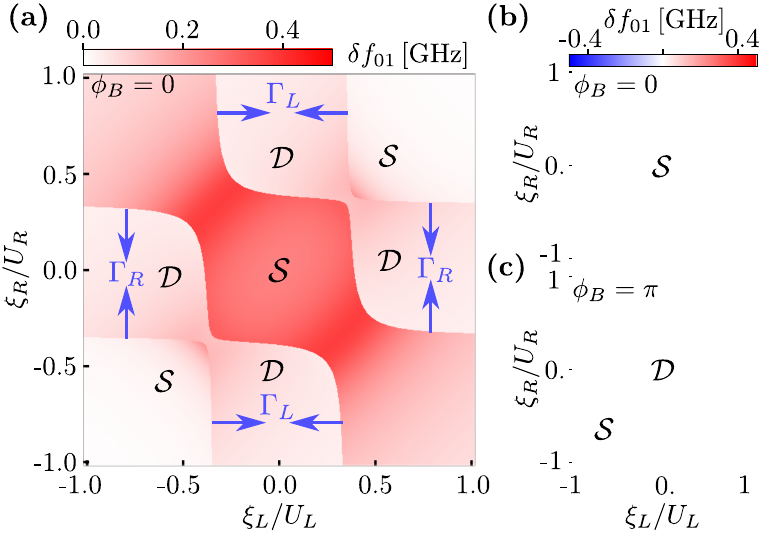} 
\caption{\label{fig:Charging}
(a) Charging diagram for the DQD junction plotted in DQD groundstate transmon detuning, for $\Gamma_{L/R}=0.75 \Delta$ and $\phi_B = 0$. Letters $\mathcal{S}$, $\mathcal{D}$ indicate groundstate parity, and blue arrows indicate the closing of doublet sectors for increasing $\Gamma_{L/R}$. (b, c) Charging diagram at $\Gamma_{L/R} = 1.2 \Delta$ for (b) $\phi_B = 0$ and (c) $\phi_B = \pi$. All plots use $t_d = 0.25\Delta$.  
} 
\end{figure}

\section{System Characterization}\label{sec:Characterization}
A fundamental challenge for characterizing quantum dot structures in Josephson junctions is the relatively weak CPR~\cite{Hermansen2022Feb, Kurilovich2021Nov}, resulting in a small critical current and Josephson energy. This stems from the fact that for a Cooper-pair to coherently cross the junction it has to cotunnel across dots with large charging energies. Consequently, previous DQD junctions report critical currents on the scale $I_c\sim0.1 - 1.0$~nA corresponding to $V_0\sim0.1 - 1.0$~GHz~\cite{EstradaSaldana2018Dec, Bouman2020Dec}. This brings into question if it is feasible to characterize and operate a DQD with cQED as the first term of Eq.~(\ref{eq:Detuning}), corresponding to the \textit{adiabatic response}~\cite{Park2020Aug}, is proportional to $V_0$ for a sinusoidal CPR. Recent Andreev spin-qubits, however, face similar challenges, as the phase dependent spin-splitting from spin-orbit coupling falls within the same range of $\sim0.1 - 1.0$~GHz, which was successfully resolved via two-tone measurement~\cite{Hays2021Jul, Pita-Vidal2023May, Bargerbos2023Aug}.

In Fig.~\ref{fig:Charging} (a) we show a charging diagram for the two dots, with sectors visible via groundstate transmon detuning, $\delta f_{01}$ obtained from Eq.~(\ref{eq:Detuning}), and for intermediate coupling to the superconducting leads, $\Gamma_L = \Gamma_R = 0.75 \Delta$ and phase $\phi_B = 0.0$. In this regime the groundstates are primarily defined by simple filling of the DQD with each dot going from $2$ to $0$ occupancy as $\xi_{L/R}/U_{L/R}$ goes from $-1$ to $1$ \cite{Grove-Rasmussen2018Jun, EstradaSaldana2018Dec}. Each odd parity state is a spin-degenerate doublet, denoted by $\mathcal{D}$, while each even parity state is a singlet, labelled $\mathcal{S}$. At the center, $\xi_L \approx \xi_R \approx 0.0$, both dots are at half-filling and the groundstate is the exchange singlet, $\mathcal{S}_{ex}$. For our parameters, all terms of Eq.~(\ref{eq:Detuning}) provide detuning on similar scale, and the increase of $\delta f_{01}$ towards the center stems both from the stronger CPR and the enhanced matrix elements $M_{0n}$, due to the $t_d$ hybridized nature of the groundstate.

As the coupling to leads increases the doublet sectors shrinks, as screened singlet states with a quasiparticle bound to the half-filled dot replaces the spin-half odd parity state. 
In Fig.~\ref{fig:Charging} (b, c) we show the charge diagram at $\Gamma_{L} = \Gamma_{R} = 1.2 \Delta$, just after the closing of the $\mathcal{D}$ sectors. Here, the structure of dot charging has vanished and instead, for $\phi_B = 0$, only a peak of transmon detuning is visible. At this coupling, singlet groundstates of all even fillings are adiabatically connected by both $t_d$ and $\Gamma_{L/R}$, with no quantum phase transitions in between \cite{EstradaSaldana2020Nov}. For $\phi_B = \pi$ a doublet island emerges at the center with a positive detuning, while the immediate surrounding singlet state has a negative one. The origin of these states are explained in the next section. The detuning signs relate to the $\pi$-phase ($0$-phase) of the doublet (singlet) groundstate~\cite{Bouman2020Dec}, rendering the adiabatic term of Eq.~(\ref{eq:Detuning}) positive (negative). The amplitude of $\delta f_{01}$, in conjunction with the groundstate dependent charge diagrams, highlights that characterization of a DQD is feasible with two-tone spectroscopy. 

\subsection{(1,1)-Sector}

\begin{figure*}[t]
\includegraphics[width=0.85\linewidth]{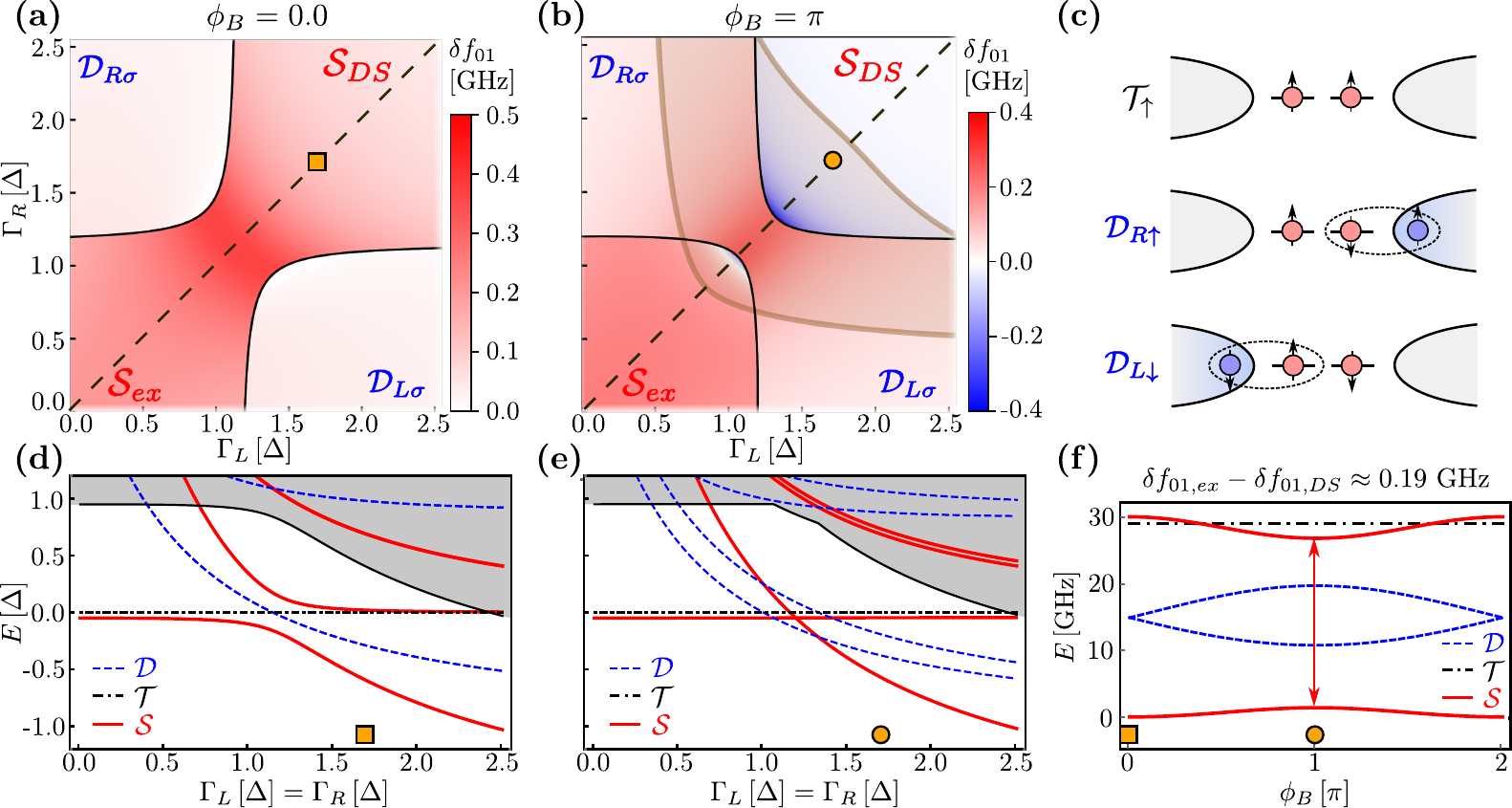}
\caption{\label{fig:11Sector}  $\mid$
(a, b) Transmon detuning as a function of couplings $\Gamma_L$ and $\Gamma_R$ at the particle-hole symmetrical point, $\xi_L = \xi_R = 0.0$, for $\phi_B = 0$ and $\phi_B = \pi$ respectively. Black contours indicate parity transitions, and the brown overlay in (b) indicate the area for which $2|E_{DS}-E_{ex}| < \Delta + \min{E_D} - E_g$ relevant for qubit operation. (c) Sketch of the triplet state $\mathcal{T}_\uparrow$ and screened doublet states $\mathcal{D}_{R\uparrow}$ and $\mathcal{D}_{L\downarrow}$. (d, e) Linecuts of (a) and (b) respectively, indicated by dashed-lines, highlighting the control of $\phi_B$ over anti-crossings between singlet and doublet states. Grey overlay indicates expected position of the superconducting continuum, given by $E_g + \Delta$, which is not captured by the MGAL description. (f) Phase evolution at $\Gamma_L = \Gamma_R = 1.7 \Delta$ showcasing an operational point for the bond qubit. The red line indicates the qubit transisiton at the optimal $\phi_B = \pi$ point, with distinguishing transmon detuning shown above. Orange symbols indicate parameters in relation to previous plots. All plots use $t_d = 0.25\Delta$.
}
\end{figure*}

Precisely at the particle-hole symmetrical point, $\xi_L = \xi_R = 0.0$, the screening quasiparticles are equal superpositions of electron and hole and so effectively chargeless. This fact, combined with the large charging energy, makes the DQD states at the $(1,1)$ sector protected from charge noise to linear order. Consequently, we will focus on this gating in the remaining sections. As to obtain a minimal model for the $(1,1)$-sector we perform Schrieffer-Wolff transformations of Eq.~(\ref{eq:FullHam}) valid for $\tilde{U}-|\tilde{\Gamma}_L - \tilde{\Gamma}_R| \gg \tilde{t}_d$, for both the $\mathcal{S}$ and $\mathcal{D}$ parity sectors. Keeping only terms up to linear order in $\tilde{\xi}_{L/R}/\tilde{\Gamma}_{L/R}$ the effective Hamiltonian becomes, 
\begin{align}
&H \approx H_{\mathcal{S}}+\sum_\sigma H_{\sigma} + H_T, \label{eq:MinModel} \\
&H_{\mathcal{S}} = \begin{pmatrix}
-\tilde{\Gamma}_L - \tilde{\Gamma}_R && t_+^* \\
t_+ && -\tilde{U} - E_{ex} 
\end{pmatrix}
\begin{matrix}
\ket{\mathcal{S}_{DS}} \\
\ket{\mathcal{S}_{ex}}
\end{matrix}, \label{eq:SHam} \\
&H_{\sigma} = \begin{pmatrix}
-\tilde{U}/2-\tilde{\Gamma}_L - E_D && t_-^*/\sqrt{2} \\
t_-/\sqrt{2} && -\tilde{U}/2-\tilde{\Gamma}_R  - E_D
\end{pmatrix}
\begin{matrix}
\ket{\mathcal{D}_{L\sigma}} \\
\ket{\mathcal{D}_{R\sigma}}
\end{matrix},
\end{align}
with states to the right indicating different subspaces. Here, $H_T$ is composed of three decoupled spin-$1$ triplet states with energy $E_T = -\tilde{U}$. The singlet and doublet exchange energy is given by,
\begin{align}
E_{ex} &= \frac{|t_+|^2}{\tilde{U}+\tilde{\Gamma}_L+\tilde{\Gamma}_R} + \frac{2\tilde{U}|t_-|^2}{\tilde{U}^2 - (\tilde{\Gamma}_L-\tilde{\Gamma}_R)^2}, \\
E_D &= \frac{|t_+|^2}{\Gamma_L + \Gamma_R}
\end{align}
which separates $\ket{\mathcal{S}_{ex}}$ from the triplet states, $\ket{\mathcal{T}_\sigma}$, and $t_\pm$ are effective couplings defined by
\begin{align}
t_+ &= \sqrt{2}\tilde{t}_d \cos{\frac{\phi_B}{2}} + i\frac{\tilde{t}_d}{\sqrt{2}}\left(\frac{\tilde{\xi}_L}{\tilde{\Gamma}_L} - \frac{\tilde{\xi}_R}{\tilde{\Gamma}_R}\right) \sin{\frac{\phi_B}{2}}, \label{eq:tp} \\
t_- &= -i \sqrt{2} \tilde{t}_d \sin{\frac{\phi_B}{2}} - \frac{\tilde{t}_d}{\sqrt{2}}\left(\frac{\tilde{\xi}_L}{\tilde{\Gamma}_L} + \frac{\tilde{\xi}_R}{\tilde{\Gamma}_R}\right) \cos{\frac{\phi_B}{2}}.
\end{align}
The doublet parity states, $\ket{\mathcal{D}_{j\sigma}}$, correspond to partially screened states with one dot screened, indicated by index $j$, and the other dot with a free spin $\sigma$, e.g. $\ket{D_{L\uparrow}} = \ket{\mathcal{S}_{L}}\otimes \ket{\uparrow_R}$. A derivation of this Hamiltonian can be found in the supplemental material~\cite{SM}.

In Fig.~\ref{fig:11Sector} (a, b) we show the groundstate transmon detuning as a function of couplings $\Gamma_{L/R}$ at the particle-hole symmetrical point, $\xi_L = \xi_R = 0.0$, and in (c) an illustration of the $T_\sigma$ and $D_{j\sigma}$ states. The observed behaviour can be understood in terms of Eq.~(\ref{eq:MinModel}); at low couplings, $\Gamma_L, \Gamma_R \ll U$, the exchange singlet, $\mathcal{S}_{ex}$, is the groundstate and as $\Gamma_{j}$ is increased a parity transition into $\mathcal{D}_{j\sigma}$ occurs around $\tilde{\Gamma}_j \approx \tilde{U}/2$. However, for symmetrical increase of couplings, $\Gamma_L = \Gamma_R$, the state $\mathcal{S}_{ex}$ continuously evolves into the double screened singlet, $\mathcal{S}_{DS}$, via an anti-crossing. Close to this anti-crossing, $\tilde{\Gamma}_L = \tilde{\Gamma}_R \approx \tilde{U}/2$, the groundstate is determined by $t_+$ and $t_-$ which governs the size of the anti-crossing of singlets and doublets respectively. By tuning the phase to $\phi_B = 0.0$ ($\phi_B = \pi$) the coupling $t_+ = 0$ ($t_- = 0$) is removed and an anti-bonding doublet (singlet) state becomes the groundstate. This behavior is shown in Fig.~\ref{fig:11Sector} (d, e) where for $\phi_B = 0$ the groundstate is always singlet, while for $\phi_B = \pi$ the energy gain from doublet anti-bonding creates a window where the groundstate is doublet. 

For qubit operation between $\ket{\mathcal{S}_{ex}}$ and $\ket{\mathcal{S}_{DS}}$ the most promising tuning is either $\phi_B = 0$ or $\phi_B=\pi$ which both maximises the state distinguishing signal, $\delta f_{01,ex} - \delta f_{01,DS}$, and corresponds to a sweet spot in $\phi_B$, resulting in quadratic protection to fluctuations. Furthermore, as in our scheme we wish to drive transistions via dot gates, controlling $\xi_{L/R}$, we focus on $\phi_B = \pi$. Here, the the two singlet states are decoupled as $\xi_L = \xi_R = 0.0$, but feels linear perturbations of $\xi_{L/R}$ maximally, as seen in Eq.~(\ref{eq:tp}). Lastly, as to avoid incidental driving into the DQD doublet parity sector we require $2|E_{ex} - E_{DS}| < \Delta + \min{E_{D}}-E_g$ with $\min{E_D}$ indicating the lowest energy of any doublet parity state and $E_g$ groundstate energy. If this criterion is not satisfied the driving of the transition $\ket{\mathcal{S}_{ex}}\leftrightarrow \ket{\mathcal{S}_{DS}}$ also 
facilitates $\ket{\mathcal{S}_{DS/ex}} \rightarrow \ket{\mathcal{D}}\otimes \ket{qp}$ with the latter indicating the lowest energy doublet state in addition to a free quasiparticle in the continuum, thus driving one out of the operational basis. The area of $\Gamma_{L/R}$ space supporting this restriction is shown in Fig.~\ref{fig:11Sector} (b) as a brown overlay. In addition, if either of the singlets are the groundstate then initialization can be performed by simply waiting for the system to equilibrate. The overlap of the brown overlay with singlet groundstate in Fig.~\ref{fig:11Sector} (b) supports this option. Else, initialization of the operational basis can be done via dot gate operations, as detailed in the supplemental~\cite{SM}, removing the groundstate restriction if required. In Fig.~\ref{fig:11Sector} (f) we plot the excitation spectrum in units of GHz at a specific realization of the above restrictions, namely $\Gamma_L = \Gamma_R = 1.7 \Delta$. with $t_d = 0.25\Delta$. At this point $\delta f_{01,ex} \approx 0.16$~GHz and $\delta f_{01,DS}\approx 0.02$~GHz, allowing one to identify the two singlets via single-shot readout of the resonator. Next, we focus on this specific realization and detail qubit gate operations.

\section{Qubit Operation}\label{sec:QubitOp}

\begin{figure}[t]
\includegraphics[width=1\linewidth]{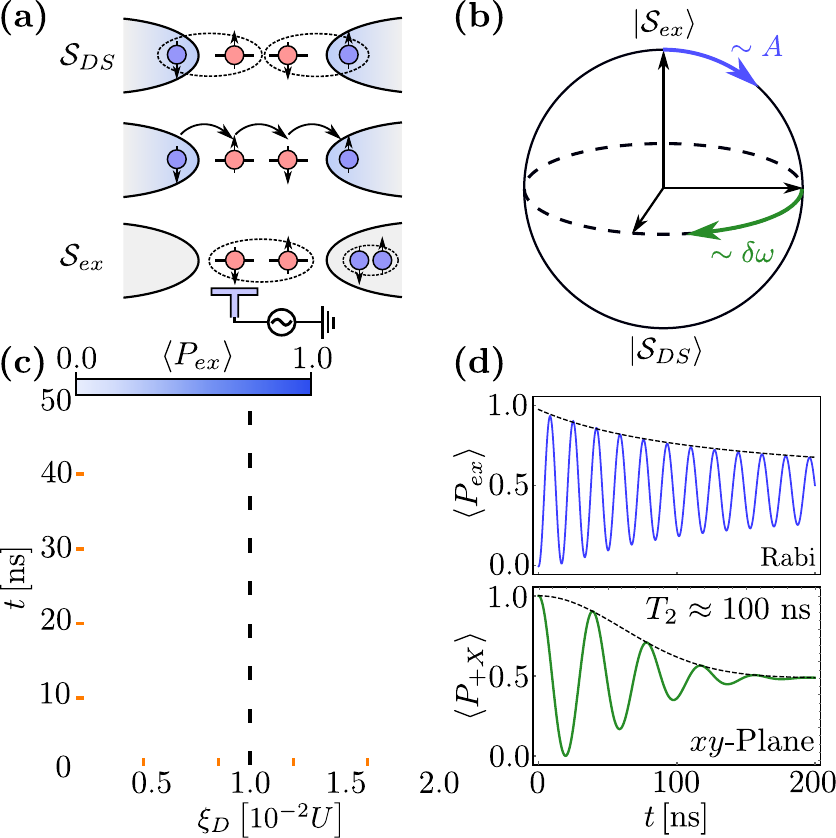}
\caption{\label{fig:QubitOperation}  $\mid$
(a) Schematic of the coupling inducing $\ket{\mathcal{S}_{DS}} \leftrightarrow \ket{\mathcal{S}_{ex}}$ transitions by driving of $\xi_L$. (b) Bloch sphere in the bond qubit basis with rotations indicated by driving term. (c) Rabi transfer probability as a function of driving time and strength, with decay due to $1/f$ noise plotted at the $\Gamma_{L}=\Gamma_R=1.7\Delta$ point with zero detuning, $f_D = f_S\approx 25$~GHz. Dashed line corresponds to plot (d). (d) Top: Rabi oscillations for $\xi_D = 10^{-2}U$ and algebraic decay due to $1/f$ noiseusing Eq~(\ref{eq:PexRabi}), fitted to $y(t) = \frac{1}{2}+\frac{1}{2}\left(1+\frac{t}{1.5 T_2}\right)^{-1}$ (dashed line). Bottom: Rotations in the $xy$-plane starting from $\ket{+X} = \frac{1}{\sqrt{2}}\left(\ket{\mathcal{S}_{DS}}+\ket{\mathcal{S}_{ex}}\right)$ with detuning $\delta f = 10^{-3} \Delta/h$ using Eq.~(\ref{eq:P+xPlane}), and exponential decay due to $1/f$ noise fitted to $y(t)=\frac{1}{2}+\frac{1}{2}\exp\left[-\frac{t^2}{T_2^2}\right]$ (dashed line).   
}

\end{figure}

To facilitate qubit operations in the bond qubit basis we can drive $\xi_L$ and/or $\xi_R$, controlled by respective dot gates. In the following we set $\tilde{\xi}_R = 0$ and $\tilde{\xi}_L = \tilde{\xi}_{D}\sin\left(2\pi f_D t\right)$, amounting to left dot driving at the particle-hole symmetrical point. Additionally assuming $\phi_B = \pi$ the singlet Hamiltonian, Eq.~(\ref{eq:SHam}), takes the form of a stereotypical qubit Hamiltonian. Then, by gauge transforming with $\mathcal{U}=\exp\left[-i\pi f_D t \sigma_z\right]$, with $t$ denoting real time and $\sigma_j$ Pauli matrices, and by performing a rotating-wave approximation we arrive at the following Hamiltonian,
\begin{align}
&\hat{H}_{\mathcal{S}} = h\frac{f_S-f_D}{2}\sigma_z + \frac{A}{2} \sigma_x, \\
&A = \frac{\tilde{t}_d}{\sqrt{2}}\frac{\tilde{\xi}_D}{\tilde{\Gamma}_L}, \nonumber \\
&hf_S = \tilde{\Gamma}_L + \tilde{\Gamma}_R - \frac{\tilde{U}}{2} - E_{ex}, \nonumber
\end{align}
with $f_S$ denoting the bond qubit level splitting, and $A$ the Rabi coupling which can be tuned via $t_d$. The validity of this approximation requires $A/h\ll f_S+f_D$ as to discard counter rotating terms. Physically, the driving of $\xi_D$ amounts to momentarily reintroducing a $\sin{\phi_B/2}$ term into the CPR which at $\phi_B = \pi$ corresponds to driving a supercurrent back and forth. If this driving is stopped at an appropriate time, given by $A$, in total a single chargeless quasiparticle has crossed the junction and the system has transitioned from $\ket{\mathcal{S}_{DS}}$ to $\ket{\mathcal{S}_{ex}}$ as depicted in Fig.~\ref{fig:QubitOperation} (a). 
To perform rotations around the $x$-axis one drives $\xi_L$ with $f_D = f_S$, resulting in finite $A$ with no detuning, and therefore Rabi oscillations in the bond basis. For rotations around the $z$-axis one stops the driving, $A=0$, while keeping a finite detuning, $\delta f = f_D - f_S$, the effect of which could be measured via Ramsey spectroscopy. These operations are depicted in Fig.~\ref{fig:QubitOperation} (b) and yields full mobility across the Bloch sphere. 

\subsection{Qubit-Qubit coupling}
Coupling of the YSR bond qubit to other qubits can be done via standard cQED techniques, since the bond qubit states couple to the circuit via CPR. As an example, bringing the transmon on resonance with the qubit, $f_{0}\approx f_{S}$, yields anti-crossings with splitting given by the matrix element $M_{DS,ex}\sim 1$~GHz (Eq.~(\ref{eq:MatrixElement})) shown in the supplemental material~\cite{SM}, placing us in the strong coupling regime and thereby facilitating transmon-qubit superpositions and further superpositions between distant qubits~\cite{Pita-Vidal2023May, Pita-Vidal2024May, Pita-Vidal2024MayV2}.  

\section{Qubit Noise and Sweetspots} \label{sec:NoiseAndSweetspots}
In this section we discuss the predominant sources of noise to the YSR bond qubit. We focus on slowly varying noise associated with the zero frequency component of the noise spectrum, $S_{\chi}(f\rightarrow 0)$, for $1/f$ noise which is typically present in capacitively coupled gates and in Overhauser fields from magnetic impurities~\cite{Petta2005Sep, Kuhlmann2013Sep}. These sources act as a random initialization of system parameters for each run, and in measurements where many runs are sampled they will influence the average outcome~\cite{Ithier2005Oct}. First, we focus our attention on the couplings $\Gamma_{L/R}$ and $t_d$ which we expect, due to the bond nature of the qubit, to constitute the primary source of noise. To obtain a conservative estimate we assume that typical capacitive gates experience gaussian voltage fluctuations with variance $\sigma_V\sim 0.2$~mV \cite{Hays2021Jul, Pita-Vidal2023May}, and from Ref.~\cite{EstradaSaldana2020Nov} (See Fig.~4 and Fig.~7) we find an approximate change of $\Gamma_R\in \left[0.3, 0.8\right]$~meV for a gate voltage change of $V_g\in \left[-0.5, -2.0 \right]$~V in the kind of device we are modelling. As both $\Gamma_{L/R}$ and $t_d$ describes wave-function overlap, which is controlled by gating a barrier, we assume for simplicity that they have similar noise. Further assuming that both $t_d$ and $\Gamma_{L/R}$ changes linearly as a function of gate, we get the following noise
\begin{equation}
\sigma_{\Gamma} = \sigma_{t_d} \approx \frac{d\Gamma}{dV_g} \sigma_V \approx 0.1~\mu\text{eV}
\end{equation}
These numbers also depend on the exactness of the models fitting the aforementioned experiments and neglects the exponential dependence on gate, typical of a tunnel junction, and should in that light be regarded as fairly rough estimates. Nonetheless using these numbers we find the following standard deviations on qubit parameters
\begin{align}
\sigma_{f_S} &\approx 0.07~\mu\text{eV}/h, \\
\sigma_{A} &\approx 2\cdot10^{-3} A,
\end{align}
with the noise on coupling, $A$, being proportional to the coupling itself. In the following we consider pulses up to $\xi_D \approx 10^{-2}U$, well within the limit of $\sqrt{\tilde{\Gamma}_{L/R}^2+\tilde{\xi}_D^2}\approx \tilde{\Gamma}_{L/R}$ necessary for the minimal model to be valid, which for our DQD parameters yields $\sigma_A \approx 10^{-3}~\mu$eV. Consequently, noise on the qubit frequency, $\sigma_{f_S}$ constitutes the main source of decoherence and in the following we neglect effects of $\sigma_A$. Starting from $\ket{\mathcal{S}_{DS}}$ at $t=0.0$, we find the following transition amplitude for Rabi driving with $f_S - f_D = x_{f}$ with $x_{f}$ denoting frequency fluctuations
\begin{align}
&\langle P_{ex}(t) \rangle = \frac{1}{\sqrt{2\pi}\sigma_{f_S}}\int_{-\infty}^\infty dx_{f} P_{ex}(t)\exp\left[-\frac{1}{2}\frac{x_f^2}{\sigma_{f_S}^2}\right] \label{eq:PexRabi} \\
& = \bigg\langle \frac{1}{2}\frac{A^2}{A^2 + \left(h x_{f}\right)}^2\left[1 - \cos\left(\sqrt{A^2+\left(h x_{f}\right)^2}t/\hbar\right)\right]\bigg\rangle \nonumber
\end{align}
with $P_{ex}(t) = |\braket{\mathcal{S}_{ex}(t)}|^2$. This results in an algebraic decay as a function of $t$ for our parameters. On the other hand, for rotations in the $xy$-plane with no driving, $A=0$, and finite detuning, $f_S -f_D = \delta f_0 + x_{f}$, and using initial state $\ket{+X} = \frac{1}{\sqrt{2}}\left(\ket{\mathcal{S}_{ex}}+\ket{\mathcal{S}_{DS}}\right)$ we find,
\begin{equation}
\langle P_{+X} \rangle = \frac{1}{2} + \frac{1}{2}\cos\left(2\pi f_S t\right)\exp \left[-2\pi^2 \sigma_{f_S}t^2\right], \label{eq:P+xPlane}
\end{equation}
which corresponds to an exponential decay with a decoherence time of $T_2 \approx 100$~ns. The effect of these operations are shown in Fig.~\ref{fig:QubitOperation} (c, d) demonstrating the feasibility of Rabi pulses and Ramsey measurements for our operational parameters.

In this analysis we have assumed that the charging energy, $U$, and gap, $\Delta$, are not susceptible to noise since they are material and geometry defined parameters. If present, noise on $U$ would linearly affect $f_S$, and $\Delta$ would effect both $f_S$ and $A$ through the rescallings in Eq.~(\ref{eq:Rescallings}). Additionally, in the above we have assumed that noise between different gates is uncorrelated. In realistic setups capacitive cross-talk between gates is common, which yield correlated noise between gates. Assuming that such correlation affect only neighboring gates, we find little effect as gates controlling $\Gamma_{L/R}$ and $t_d$ are separated by gates controlling $\xi_{L/R}$, which are unaffected by slow noise due to the particle-hole sweet-spot. Although the above estimates are obtained through the MGAL approximation, we consider them fairly accurate as $\sigma_{f_S}$ and $\sigma_A$ are spectral quantities, and should therefore compare to results from Numerical Renormalization Group.

\begin{figure}[h]
\includegraphics[width=1\linewidth]{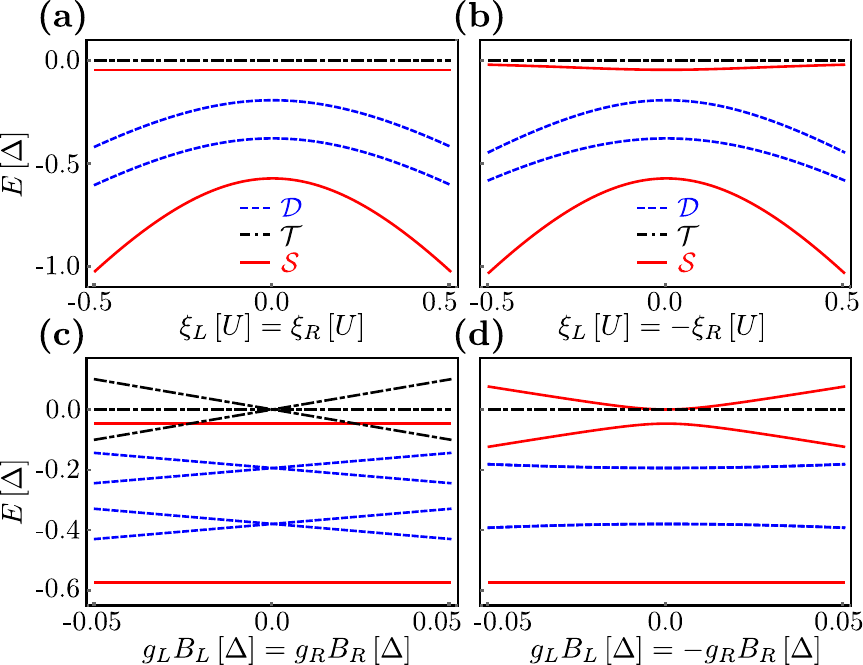}
\caption{\label{fig:SweetSpot}  $\mid$
Energy dispersion of states in the 11-sector calculated from Eq.~(\ref{eq:FullHam}) with Eq.~(\ref{eq:HamB}) included. (a, b) The quadratic protection in both symmetric and asymmetric tuning of $\xi_j$ of the singlet states is shown. (c, d) Similarly for symmetric and asymmetric fields. Note that in (d) one triplet state mixes with the exchange singlet, and is therefore depicted red although both singlet and triplet are now somewhat spin polarized.
}
\end{figure}
\subsection{Sweet Spots}
Next we discuss the protection from dephasing offered by the sweet spots in both quantum dot gates and from magnetic fields. In our original Hamiltonian, Eq.~(\ref{eq:FullHam}), we include the following term,
\begin{equation}
H_B = g_LB_L\left(n_{L\uparrow}-n_{L\downarrow}\right) + g_RB_R\left(n_{R\uparrow}-n_{R\downarrow}\right) \label{eq:HamB}
\end{equation}
with $g_{L/R}$ being dot g-factors and $B_{L/R}$ magnetic fields which may be different due to local fluctuations on each dot. For both symmetric and anti-symmetric magnetic fields and dot gate detuning from the sweet spot, we find no linear terms in the qubit energy, $f_S$, as shown in Fig.~\ref{fig:SweetSpot}. For gate detuning this protection originates from the particle-hole symmetry present at the center of the (1,1)-sector, while the protection from symmetric field, $g_LB_L = g_RB_R$, is simply due to the non-magnetic nature of the singlet basis. For asymmetric fields the exchange bond provides quadratic protection since any magnetic polarization of the electrons would diminish the gain of exchange energy of the $\ket{\mathcal{S}_{ex}}$ state. If magnetic fluctuations becomes of the order of the exchange splitting itself, $g_jB_J \sim E_{ex}$, the protection would become insufficient and mixing of the operational basis with triplet states could occur. A possible complication present in a more rigorous treatment would be that the screening quasiparticles experience a different field than the dots, such from different material $g$-factors~\cite{Pavesic2023Aug}. Such a term would lead to a quadratic dispersion of the $\ket{\mathcal{S}_{DS}}$ state as a function of field, not present in Fig.~\ref{fig:SweetSpot} (c, d), since the polarization of quasiparticles and dot electrons would counteract the screening bond. However, similar to the interdot exchange bond, this results in protection up to the scale of the bonding energy itself, which is given by $\Gamma_j$, and so field protection is expected to remain. Lastly, it should be noted that the MGAL rescallings, Eq.~(\ref{eq:Rescallings}), used throughout this paper does not include magnetic fields, and so possible rescalling of $B_{L/R}$ with $\Gamma_{L/R}$ might be required in a realistic model~\cite{Zonda2023Mar}. In the supplemental material we compare this to Zero-bandwidth modeling and find equal magnetic protection~\cite{SM}

\subsection{Qubit Depolarization}
Another important source of noise is the depolarization that occurs due to direct transitions within the qubit basis. In the low temperature limit, $k_BT \ll h f_S$, only relaxation processes from the higher energy qubit state to the lower one is relevant. The transition rate of such processes can be estimated via Fermi's golden rule,
\begin{equation}
\Gamma_{1} = \frac{\pi}{2\hbar^2}\sum_\chi 
|\bra{\mathcal{S}_{ex}}\frac{\partial H}{\partial \chi}\ket{\mathcal{S}_{DS}}|^2S_\chi(f_S) 
\end{equation}
with $\chi$ spanning parameters of the Hamiltonian and $S_\chi(f_S)$ indicating the corresponding noise spectrum at qubit frequency. Since we a priori do not know the associated spectrum's we simply discuss possible couplings associated with the matrix element of $\frac{\partial H}{\partial \chi}$. For the proposed operational point one finds significant coupling to both phase, via $\phi_B$, and dot gate, $\xi_{L/R}$, noise as they enter directly into the off-diagonal component of Eq.~(\ref{eq:SHam}). On the other hand, noise on $t_d$ does not introduce depolarization and noise on $\Gamma_{L/R}$ does so only weakly through the rescalings of off-diagonal parameters via Eq.~(\ref{eq:Rescallings}). In an experimental setup these noise sources would have to be explored more thoroughly in order to avoid significant depolarization. By choosing another operational point of the bond qubit, namely $f_S \approx 0.0$ and $\phi_B=0.0$, one can remove the depolarization from both $\phi_B$ and $\xi_{L/R}$, but one reintroduces finite terms for $t_d$ and $\Gamma_{L/R}$ in addition to enhancing the direct dephasing. This operational point is discussed further in supplemental material~\cite{SM}. The matrix elements discussed above depends on the state composition of the qubit states, and results from MGAL are therefore not completely accurate. To account for this, we compare MGAL results to a Zero-bandwidth model in the supplemental material~\cite{SM}, which yields similar matrix elements.

Lastly, we comment on quasiparitcle poisoning which corresponds to processes where an excited quasiparticle, with energy $E_{qp}\approx \Delta$, induces a transition from state $n$ to $m$ emitting energy $E_{qp}-E_n+E_m$~\cite{Hays2018Jul, Hays2020Nov}. These processes are parity changing and so brings the qubit out of the computational basis. However, in quantum dots poisoning processes are not observed in DC transport~\cite{Steffensen2022Apr} except at elevated temperatures~\cite{Kumar2014Feb}, and using cQED techniques the parity transition rate was estimated to be on the scale of $\sim 0.1-1.0$~ms~\cite{Bargerbos2022Jul}, much slower than the estimated $T_2$ time. Consequently, we do not consider these processes to be important for YSR bond qubit operation.

\subsection{Additional Terms}
Other terms not included in the investigated Hamiltonian might also influence the qubit performance. Here we discuss two terms present in realistic devices, namely interdot charging, $U_{LR}$, and spin-orbit coupling. For small to moderate interdot charging energy, $U_{LR} \lesssim U$, we do not expect any qualitative change to the discussions above, as the particle-hole symmetrical point would simply be displaced, but would still provide the quadratic protection in $\xi_{L/R}$. A $U_{LR}$ term, however, might quantitatively affect both the optimal operational points and the measured transmon detunings, $f_{01,n}$. A more detailed analysis of this effect has been neglected for the sake of simplicity.

Spin-orbit coupling is typically modelled as spin-flipping hopping terms in DQD's, and for finite magnetic field these induces couplings between singlet states and triplet states. For zero-magnetic field and uni-directional spin-flipping no such coupling is induced~\cite{Bouman2020Dec}. If multiple dot orbitals are considered spin-orbit might also introduce spin-spin interactions in the system, as utilized for operation in Ref.~\cite{Pita-Vidal2023May}. The presence of either spin-spin interactions or singlet-triplet couplings, if on the order of the exchange coupling, $E_{ex}$, would serve as source of depolarization driving the qubit out of its operational basis. In general however, for small spin-orbit coupling compared to normal tunneling we expect no qualitative changes to the proposal above. Lastly, we emphasize that one of the attractions of our geometry is that significant spin-orbit coupling is not required in the material platform. This allows e.g. Si based quantum dots to be employed, which can be grown via CMOS techniques~\cite{Xue2021May}

\section{Discussion and Conclusion} \label{sec:Discussion} 
\subsection{Qubit Comparisons}

In this section we compare and relate the YSR bond qubit to other solid-state qubit designs.

\textit{The spin qubit}, based on the electron spin of quantum dots~\cite{Burkard2023Jun}, remain one of the leading qubit platforms. These devices offers small device blueprints, compared to the $\mu$m scale of typical superconducting qubits, and do not require superconductivity, thus allowing for higher temperature operation. Typical designs often employ multiple QDs, such as the triplet-singlet qubit~\cite{Petta2005Sep}, as to facilitate qubit initialization and operations through electrostatic gates, replacing the need for control of local magnetic fields~\cite{Loss1998Jan}. Here, the singlet state is an exchange bonding state, similar to $\ket{\mathcal{S}_{ex}}$ in our proposal, while the other qubit state is a magnetic triplet state. This causes the qubit to be sensitive to local magnetic noise, which influences the qubit frequency, in addition to requiring either slow Overhauser fields~\cite{Petta2005Sep}, micromagnets~\cite{Wu2014Aug}, or materials with strong spin-orbit coupling~\cite{Jirovec2021Aug} to facilitate qubit operation. Designs such as the exchange-only singlet-only qubit~\cite{Sala2017Jun} obtain protection from local magnetic fields through the use of singlet-singlet qubit basis, similar to the YSR bond qubit, but require more QDs and gates to operate. Such exchange-only designs, however, couple only locally through electron tunneling, rendering long range qubit correlations hard to achieve.

\textit{Superconducting qubits}, of which transmon~\cite{Koch2007Oct} and fluxonium~\cite{Manucharyan2009Oct} are the leading platforms, offers long range qubit correlations and access to qubit gates through use of cQED technique. However, the large $\mu$m device blueprints limits the number of qubits that can be placed on a single chip. This limitation inspired the development of \textit{Andreev qubits}, which attempt to combine the advantageous of spin and superconducting qubits, i.e. small size and cQED capabilities. This is also the goal of the YSR bond-qubit, and it can thus be considered a kind of Andreev qubit. The original designs used cQED to manipulate local quantum states in quantum point contacts~\cite{Janvier2015Sep, Hays2021Jul, Matute-Canadas2024May}, however the QD based Andreev spin qubit~\cite{Pita-Vidal2023May, Pavesic2024Feb} relies instead on QDs to host its qubit states. This grants both charge noise protection and greater tunability, and can be utilized to make the groundstate a qubit state for easy initialization. The use of a magnetic basis, however, reinstates many issues from spin-qubits, such as sensitivity to Overhauser fields and requirement of spin-orbit and Zeeman splitting~\cite{Pita-Vidal2023May}. In this manner, the YSR bond qubit can be considered a kind of singlet-only Andreev spin qubit using YSR exchange bonds instead of additional QDs, as in Ref.~\cite{Sala2017Jun}, to provide magnetic protection and access to cQED operation. 

Our proposal is not the first to utilize YSR interactions in forming the computational subspace. Other designs are either relying on the interactions of classical spins on the surface of substrates~\cite{Mishra2021Dec}, or the discreteness of a Richardson superconductor~\cite{Pavesic2022Feb}. This latter proposal also utilizes a singlet-based computational subspace, which at sweets-spots is insensitive to charge noise. Our proposal further integrates cQED techniques to operate and manipulate rendering it into an Andreev qubit, in addition to not relying on asymmetric superconducting leads with finite charging.

Another interesting comparison is the recent proposal of a \textit{fermion-parity qubit}~\cite{Geier2024Jun}, where the operational subspace is composed of charge-equal states of different parities, hence a 'chargeless' charge qubit.Similar to our design this qubit relies on a DQD coupled to two superconducting leads. Indeed, by characterising the left and right QD-superconductor subsystem by parity in our propasal, the transition $\ket{\mathcal{S}_{ex}}\leftrightarrow\ket{\mathcal{S}_{DS}}$ corresponds to an \textit{odd-odd} to \textit{even-even} parity transition by the tunneling of a chargeless quasiparticle from QD to QD. However, the fermion-parity qubits subspace is magnetic and so sensitive to fluctuations, and requires magnetic field to lift spin degeneracy similar to QD based Andreev spin qubit designs.

Finally, we would like to highlight the recent interest into \textit{artificial Kitaev chains} based on quantum dots~\cite{Dvir2023Feb, Bordin2024Feb}. Here, localized Andreev states and spin-orbit effect are used to engineer couplings between QDs to obtain a topological phase with localized Majorana fermions at the end sites. The Majorana's can then be used to make a topological parity qubit~\cite{Leijnse2012Oct}, which share certain properties with our design such as protection from QD gate noise ($\xi_j$), but sensitivity to noise on tunnel couplings. Such designs, however, are sensitive to correlated noise of tunnel couplings and gate (opening the sweet-spot), and correlated gate-gate noise ($\xi_L=\xi_R$). The unique possibility of non-Abelian qubit operations in larger devices, such as braiding of multiple Majorana's, distinguishes this kind of design from ours, which relies on cQED instead. We highlight that current platforms for artificial Kitaev chains now integrate YSR interactions~\cite{Zatelli2024Sep, Liu2024Jul}, and multiple proposals seek to integrate double quantum dots with cQED tunability~\cite{Samuelson2024Jan, Pino2024Feb, Tsintzis2024Feb}. Further advances in this direction would as a bi-product create promising platforms for realizing the YSR bond qubit, paving the road for an experimental demonstration.

\subsection{Conclusion}
We have explored the possibility of constructing and operating a YSR bond qubit using realistic parameters from existing experimental platform, and confirmed that this setup is characterizable via cQED measurements, and that there exists an extended parameter region where qubit operation is feasible. Furthermore, we derived an effective qubit Hamiltonian for this region and detailed possible qubit operations via gate driving. Conservative estimations of noise yields an expected $T_2$ lifetime of $~100$~ns which is sufficient for multiple Rabi oscillations due to the strong drive coupling. We highlight the quadratic protection in both quantum dot gates and magnetic field fluctuations, allowing the qubit to be fabricated in materials without spin-orbit coupling and where magnetic Overhauser noise can be present, thus greatly expanding the range of available platforms. Lastly, we bring to attention that the utilized DQD parameters and estimation of noise is based on a device not optimized for YSR bond qubit operation, and that limiting the effect of charge-noise on tunnel couplings could significantly increase $T_2$. This demonstrates the feasibility of achieving a singlet based Andreev qubit in realistic material platforms.

\newpage

\begin{acknowledgments}
We wish to thank Rubén Seoane Souto, Francisco Jesus Matute Fernandez-Cañadas, Ramon Aguado, Jens Paaske and Virgil Baran for useful comments and discussion. G. S. and A. L. Y. acknowledge
financial support from the Spanish Ministry of Science through Grant TED2021-130292B-C43 funded by
MCIN/AEI/10.13039/501100011033, ”ERDF A way of
making Europe” and the EU through FET-Open project.
AndQC.  
\end{acknowledgments} 

\end{document}